
\magnification=1200
\parskip=\medskipamount \overfullrule=0pt
\font \titlefont=cmr10 scaled \magstep3
\font \namefont=cmr10 scaled \magstep1

\def\singlespace{\baselineskip=\normalbaselineskip}

\newcount\firstpageno \firstpageno=2
\footline={\ifnum\pageno<\firstpageno{\hfil}\else{\hfil
                                                  \rm\folio\hfil}\fi}
\def\spose#1{\hbox to 0pt{#1\hss}}
\def\simlt{\mathrel{\spose{\lower 3pt\hbox{$\mathchar"218$}}
     \raise 2.0pt\hbox{$\mathchar"13C$}}}
\def\simgt{\mathrel{\spose{\lower 3pt\hbox{$\mathchar"218$}}
     \raise 2.0pt\hbox{$\mathchar"13E$}}}
\def\frac#1/#2{\leavevmode\kern.1em
 \raise.5ex\hbox{\the\scriptfont0 #1}\kern-.1em
 /\kern-.15em\lower.25ex\hbox{\the\scriptfont0 #2}}
\def\degrees{\hbox{${}^\circ$\hskip-3pt .}}
\def\pp{\par\hangindent=.125truein \hangafter=1}
\def\aref#1;#2;#3;#4{\pp #1, {\it #2}, {\bf #3}, #4}
\def\abook#1;#2;#3{\pp #1, {\it #2}, #3}
\def\arep#1;#2;#3{\pp #1, #2, #3}
\def\eq{Eq.$\,$}
\def\tT{\widetilde{T}}
\def\Ob{\Omega_{\rm\scriptscriptstyle B}}
\def\Cth{C_{\rm th}}
\def\Cthij{C_{{\rm th,}ij}}
\def\mk{\,\mu{\rm K}}
\def\ihat{{\hat\imath}}

\singlespace
\rightline{CfPA--93--TH--27}
\rightline{astro-ph/9309006}
\rightline{September 1, 1993}

\vskip 3pt plus 0.3fill
\centerline{\titlefont Implications of MAX for CDM}

\vskip 3pt plus 0.3fill
\centerline{{\namefont Mark Srednicki}}

\vskip 3pt plus 0.1fill
\centerline{Department of Physics,}
\centerline{University of California, Santa Barbara, CA 93106}

\vskip 3pt plus 0.3fill
\centerline{{\namefont Martin White, Douglas Scott and Emory T.~Bunn}}

\vskip 3pt plus 0.1fill
\centerline{Center for Particle Astrophysics}
\centerline{and Departments of Astronomy and Physics,}
\centerline{University of California, Berkeley, CA 94720}

\vskip 3pt plus 0.3fill
\centerline{ABSTRACT}
\baselineskip=15pt

We analyze the Gamma Ursae Minoris (GUM) and Mu Pegasii (MuP) scans of the
Millimeter-wave Anisotropy eXperiment in the context of cold dark matter (CDM)
models of structure formation, paying particular attention to the
two-dimensional nature of the GUM scan.  If all of the structure in the
(foreground subtracted) data is attributed to cosmic microwave background
anisotropy, then there is a detection in each scan.  For a standard CDM model,
the amplitudes of the signals are individually compatible with the COsmic
Background Explorer measurement, but marginally inconsistent with each other.

\vskip 4pt


\vfill\eject
\baselineskip=16pt

\vskip0.2in
\vskip\parskip
\noindent{\bf 1. Introduction}
\vskip0.1in

There have recently been many results quoted for anisotropies
in the cosmic microwave background (CMB) on degree scales.
Of these, the detection of a large and significant signal in the
Gamma Ursae Minoris (GUM) scan by the Millimeter-wave Anisotropy
eXperiment (MAX) [1] (see also [2]) and the strong upper limit placed
by the same experiment in their Mu Pegasii (MuP) scan [3] are the
most puzzling.

In this Letter we examine in detail the implications of this data for
cold dark matter (CDM) models of structure formation,
paying particular attention to the two-dimensional nature of
the GUM scan in our analysis.
We find that the conventional approach to calculating theoretical
predictions must be substantially modified when the scan (or scans) to be
modelled cover a two-dimensional patch of the sky.
The important point is that scanning and chopping introduce a fundamental
asymmetry into the two-dimensional autocorrelation function for the data
points.

The MAX experiment provides a good test of CDM models, since its filter
function is well-matched to the Doppler peak of the (radiation) power
spectrum (see e.g.~[4]).
Our results indicate that if all of the structure in the GUM scan is
attributed to CMB fluctuations, the results are consistent with the
normalization measured by the COsmic Background Explorer satellite
(COBE)~[5,6].  However, they are marginally inconsistent with the results from
the MuP scan.
It is possible that some of the GUM signal is not primordial, and/or that some
CMB signal was removed from MuP data by the foreground subtraction procedure;
there is no way to tell from the present data alone.
This leads us to speculate that all these results can be accommodated within
the standard CDM theory.

\vskip0.2in
\vskip\parskip
\noindent{\bf 2. Modelling MAX-GUM}
\vskip0.1in

The GUM data set consists of 165 temperatures, from the co-added
$6\,{\rm cm}^{-1}$ and $9\,{\rm cm}^{-1}$ wavebands, in a ``bow-tie''
pattern of bins covering a patch of the sky roughly
$6^{\circ}\times 1^{\circ}$ near Gamma Ursae Minoris.
During each flight the MAX telescope is scanned back and forth, tracking on
GUM, while taking data in several frequency bands.
The beam is ``chopped'' parallel to the scan direction at a frequency of
$\nu=6\,$Hz to define the temperature ``difference'' assigned
to each point.  The bow-tie pattern arises due to the effect of sky rotation
during the flight.
The scan pattern is shown in Fig.~1, in a coordinate system in which GUM
is at the origin.  At each point, the scan and chop directions are toward
(for $y<0$) or away (for $y>0$) from the origin.

In order to compare theory and experiment it is necessary to calculate the
predicted temperature autocorrelation function $\Cth$ for the scan.
To begin let $T(\theta,\phi)$ denote the temperature at a point on the sky:
$$ T(\theta,\phi)=Q\,\sum_{\ell m}a_{\ell m}\, Y_{\ell m}(\theta,\phi)
\, , \eqno(1)$$
where the $a_{\ell m}$'s are random variables whose distribution must be
specified by a model and
$Q\equiv\left\langle Q^2_{\rm RMS}\right\rangle{}^{\!0.5}
=Q_{\rm rms-PS}$~[5,6] defines the overall normalization.
In general, rotational invariance implies that
$$\langle  a^{\vphantom{*}}_{\ell m} a^*_{\ell'm'}\rangle =
 C_\ell\,\delta_{\ell\ell'}\,\delta_{mm'}\, , \eqno(2)$$
where the angular brackets denote an ensemble average over the probability
distribution for the $a_{\ell m}$'s, and $C_\ell $ is normalized so that
$C_2=4\pi/5$.
For a pure Sachs--Wolfe, $n=1$ spectrum, $C_\ell^{-1}\propto \ell(\ell+1)$.
We compute the $C_\ell$'s for CDM models using power spectra provided by
Sugiyama (e.g.~[7]), which are essentially identical with those computed by
Bond and Efstathiou (e.g.~[8]).  We restrict our study to standard CDM models
with $\Omega_0=1$ and $H_0=50\,{\rm km}\,{\rm s}^{-1}{\rm Mpc}^{-1}$, but
allow a range of values for $\Ob$.

It has become conventional in degree-scale experiments to express the
sensitivity of the experiment to the underlying power spectrum in terms of
a ``window'' or ``filter'' function $W_{\ell}$.  The autocorrelation function
is then a sum of the $C_\ell$ weighted with the $W_{\ell}$ (see e.g.~[9,4]).
This approach is perfectly adequate for experiments in which the data are
taken along a single linear scan, such as the SP91 data of Gaier et al.~[10]
or the MuP scan of MAX.  However, it is not well suited to analyzing
experiments in which the data span two dimensions, such as the GUM scan of MAX
or multiple scans of SP91.
The problem is that the chopping strategies used by these experiments define
a position-dependent, preferred direction in the sky plane.
Points which are separated by a vector parallel to the chop direction are
generally more correlated than those which are separated by a vector
perpendicular to it; furthermore these stronger correlations are negative
when the separation angle is near the peak-to-peak chop angle~[11].
This results in a strongly anisotropic autocorrelation matrix.  This is
illustrated in Fig.~2 where we show the autocorrelation function on the
sky [as given by \eq(7) below] for an ideal experiment in which the chop
direction is held fixed (parallel to the $y$-axis) for simplicity.

The anisotropy makes it difficult to derive an analytic expression for the
autocorrelation matrix, but it is easy to compute numerically.  To proceed
we define the beam-smoothed temperature in the direction $\hat{n}$ as
$$ \Theta(\hat{n}) = Q\sum_{\ell m} a_{\ell m} Y_{\ell m}(\hat{n})
\exp\bigl[-{\frac1/2}(\ell+\frac1/2)^2\sigma^2\bigr] \, , \eqno(3)$$
where $\sigma=0.425 \times 0\degrees5$ is the gaussian beam-width of the
MAX antenna.  The autocorrelation function for $\Theta$ is then
$$\eqalignno{
C_\sigma(\hat{n}_1,\hat{n}_2) &\equiv
\left\langle \Theta(\hat{n}_1)\Theta(\hat{n}_2)\right\rangle\cr
\noalign{\medskip}
&={Q^2\over 4\pi}\sum_{\ell=2}^{\infty} (2\ell+1)C_\ell\,
P_\ell(\hat{n}_1\!\cdot\!\hat{n}_2)
\exp\bigl[-(\ell+\frac1/2)^2\sigma^2\bigr]\, , &(4)\cr}$$
which specifies the predicted correlation between two beams instantaneously
separated by an angle $\theta=\cos^{-1}(\hat{n}_1\cdot\hat{n}_2)$
on the sky.
Technically one should also include the contribution from the $\ell=1$ term
in this sum; however, due to the chopping of the beam, ignoring it introduces
a negligible error in our final results.
Including the effects of the beam chopping, we can write the
temperature assigned by the experiment to the direction $\hat{n}$ as
$$ \tT(\hat{n},\ihat)\equiv \int_{0}^{1/\nu}dt\ \kappa(t)
\mathop\Theta\bigl(\hat{n}\cos\alpha(t)+\ihat\sin\alpha(t)\bigr)\, .\eqno(5)$$
Here $\alpha(t)=\alpha_0\sin(2\pi\nu t)$ accounts for the chopping motion
of the beam; $\alpha_0=0\degrees65$ for MAX.  Also, $\ihat$ is a unit vector
lying along the chop direction (which implies that $\hat{n}\!\cdot\!\ihat=0$),
and $\kappa(t)$ is the weighting factor for the beam chop.
For SP91, $\kappa(t)=2\nu\mathop{\rm sign}(t)$.
For MAX, $\kappa(t)=k\nu\sin(2\pi\nu t)$, where $k$ is determined as
follows~[12].  Take the temperature pattern in \eq(1) to be
$T(\theta,\phi)=T_0$ for $0<\theta<\pi/2$ (the northern hemisphere) and
$T(\theta,\phi)=0$ for $\pi/2<\theta<\pi$ (the southern hemisphere).
Then for any direction vector $\hat\rho$ in the $x$-$y$ (equatorial) plane,
\eq(5) should give $\tT(\hat\rho,\hat{z})=T_0$.
This condition results in
$$k^{-1} = {1\over\pi}\int_0^{\pi/2}dr\,\sin r\,{\rm erf}(\gamma\sin r)
=\Bigl({\gamma\over 2\pi^{1/2}}\Bigr) {}_1 F_1(1/2,2;-\gamma^2) \,, \eqno(6)$$
where $\gamma=\alpha_0/\sqrt2\sigma$, erf is the error function, and ${}_1F_1$
is the confluent hypergeometric function.
For $\sigma=0$, $k=\pi$; for the MAX values of $\sigma$ and $\alpha_0$,
$\gamma=2.16$ and $k=3.34$.

Now we can write the autocorrelation function for the MAX temperature
``differences'' as
$$ \eqalign{ Q^2 \Cth(\hat{n}_1,\ihat_1;\hat{n}_2,\ihat_2)&\equiv
\left\langle\tT(\hat{n}_1,\ihat_1)\,\tT(\hat{n}_2,\ihat_2)\right\rangle\cr
&= \int_0^{1/\nu} dt_1 \int_0^{1/\nu}dt_2\ \kappa(t_1)\,\kappa(t_2)\times\cr
&\qquad C_{\sigma}\bigl(\hat{n}_1\cos\alpha(t_1)+\ihat_1\sin\alpha(t_1),
  \hat{n}_2\cos\alpha(t_2)+\ihat_2\sin\alpha(t_2)\bigr)\, . \cr} \eqno(7)$$
This integral has several symmetries which can be exploited when numerically
evaluating the $165\times 165$ autocorrelation matrix, $\Cthij$, for the
GUM scan.

\vskip0.2in
\vskip\parskip
\noindent{\bf 3. Maximum Likelihood Analysis}
\vskip0.1in
\nobreak

We now turn to the limits which can be placed on $Q$, the normalization of
the fluctuation spectrum.  In accord with standard CDM models we consider
underlying cosmological fluctuations with a gaussian probability distribution.
We also assume that the experimental errors $\sigma_i$ for each $\tT_i$ are
uncorrelated and gaussian distributed.  The 165-point data set is binned
finely enough that correlations introduced by the binning process should
not be important.  The unnormalized likelihood function for $Q$ is then
given by
$$ {\cal L}(Q) \propto {1\over\sqrt{\det K}}
           \exp\bigl[-{\frac1/2}\,\tT_i(K^{-1})_{ij}\tT_j\bigr]\, ,\eqno(8)$$
where the matrix $K$ is
$$K_{ij}=Q^2\,\Cthij + \sigma_i^2\,\delta_{ij}\, . \eqno(9)$$

\eq(8) assumes that the temperatures have no systematic errors.  In fact,
MAX has a possible systematic offset which has been removed from each of
the 11 azimuthal scans of 15 data points.  This offset removal then requires
a modification of $K$.  The simplest way to implement the constraint that the
data in each azimuthal scan have zero weighted mean is to first change to a
new basis $\tT'_a=R_{ai}\tT_i$, where $\tT'_1,\tT'_2,\ldots,\tT'_{11}$ are the
weighted means of each
of the 11 scans, and $R_{ai}$ is a matrix whose linearly independent rows are
chosen so that each of the first 11 is orthogonal to each of the last 154.
Then integrating over the 11 weighted means in \eq(8) is equivalent to
replacing \eq(8) by
$$ {\cal L}(Q) \propto {1\over\sqrt{\det M}}
    \exp\bigl[-{\frac1/2}\,\tT'_a(M^{-1})_{ab} \tT'_b\bigr]\, , \eqno(10)$$
where $M_{ab}=R_{ai}K_{ij}R^T_{jb}$, and $a$ and $b$ each run only over
the range $12,\ldots,165$; that is, over the subspace orthogonal to the
one spanned by the 11 weighted means.

Before we can use \eq(10) to find the distribution of $Q$ predicted by
the data, we must choose a prior distribution for $Q$; that is, we must
decide whether equal intervals of $\log Q$, $Q$, $Q^2$,
or some other monotonic function $f(Q)$, are equally likely a priori.
Alternatively, one could decide that we actually {\it have} some prior
information and use the COBE measurement of $Q$ to fix the prior distribution,
but we will not consider that approach in this Letter.

Since the allowed range of $Q$ is from zero to infinity, the conventional
Bayesian choice (based on scale invariance arguments~[13]) is to assume a
prior distribution $f(Q)=\log Q$.
For any data set with non-vanishing errors, $\sigma_i$, this choice causes
the likelihood ${\cal L}(Q)$ to diverge for $Q\ll \sigma_i$.
This divergence can be removed by imposing a lower limit on $Q$.
For the GUM data set we find that the final results are totally insensitive
to the lower limit imposed for $Q$ once it is less than a few micro-Kelvin
(but still non-zero).

Alternatively, one could take a maximum-likelihood approach and consider
prior distributions which are uniform in some scaling variable.
The usual choice is $f(Q)=Q$ (the ``bias'' parameter $b_{\rho}$
is proportional to $Q^{-1}$, so that $dQ\equiv db_\rho/b^2_\rho$).
However, there is no compelling reason to assume a prior distribution
that is uniform in $Q$.  It is just as natural, for example, to assume that
the prior distribution is uniform in the power spectrum normalization $Q^2$.
Note that for ``good'' data, the choice of prior should make only a small
difference, so ``prior dependence'' gives us a handle on the constraining power
of the data.  In the context of small-scale CMB experiments, and assuming
that the signal-to-noise ratio exceeds unity, we expect the data to be
``good'' when the solid angle on the sky which is covered by the experiment
is much larger than that subtended by the correlation angle of the theoretical
autocorrelation function~[14].
Comparing Figs.~1 and~2, we see that this condition is fulfilled by the
GUM data set.

In Fig.~3, we show the normalized likelihood ${\cal L}(Q)\,df/dQ$
for four different choices of $\Ob$ and for $f(Q)=\log Q$.
Additionally, for $\Ob=0.01$ we show curves for $f(Q)=Q$ and $f(Q)=Q^2$.
Clearly our expectation was correct:  the effect of the prior distribution
is not very significant, and is in fact smaller than the effect of varying
$\Ob$ in the CDM model.
In Fig.~3 we also show two lines corresponding to the COBE measurement of
$Q=17\pm5\mk$ ($1\sigma$ errors).
In Table~1, we list the upper and lower limits at the 95\% confidence level
for the four values of $\Ob$ with $f(Q)=\log Q$.
We see that the GUM lower limit is within the COBE $1\sigma$ range for
all values of $\Ob$, and reaches the COBE mean value for $\Ob\simgt 0.06$.

We also wish to compare CDM with the results of the MuP scan of the MAX
experiment~[3].  This scan consists of 21 data points taken in a line.
We have used temperatures provided by the MAX group which
correspond to the second component of a two-component fit, with the first
component consisting of emission from 18$\,$K dust~[15].
These temperatures were the ones used in~[3] to set upper limits on the
observed CMB anisotropy.
An offset has been removed from this data, and we have used an appropriately
modified autocorrelation matrix.
The resulting likelihood curves are also shown in Fig.~3, and the
corresponding 95\% confidence limits in Table~1.  Note that the small $Q$
divergence in the likelihood function for $f(Q)=\log Q$ is more pronounced
for the MuP data.
For the purposes of computing confidence levels we have imposed a lower
cutoff of $0.25\mk$ (where the likelihood function begins to rise
significantly) on $Q$.
As expected from the general considerations of~[14], the MuP
results are much more sensitive to the prior distribution than are
the GUM results.
Furthermore the MuP likelihood curves have {\it larger} widths than the GUM
curves when measured in units of the corresponding mean value of $Q$.
Thus the GUM results are more statistically robust than the MuP results.
For $f(Q)=\log Q$ or $Q$, the 95\% confidence upper limits on $Q$ from the
MuP scan are just below the 95\% confidence lower limits from the GUM scan.
For $f(Q)=Q^2$, these limits overlap.

The most striking feature of Fig.~1 is that the MuP and GUM results nicely
bracket the COBE normalization.  In fact, if we compute likelihoods for the
combined MuP+GUM data set, we find that the mean values of $Q$ always lie in
the COBE $1\sigma$ range.  We note again that the apparent inconsistency
between the MuP and GUM results, when analyzed separately, could be resolved if
there is some foreground contamination in the GUM data, and/or the
two-component fit to the MuP data removed some of the actual CMB signal.  This
point of view must be considered speculative, since there is no clear candidate
for a source of foreground contamination near GUM~[1], and since the nonzero
correlation between the two components in the fit to the MuP data could mean
that part of the second component is in fact due to dust emission rather than
CMB fluctuations~[3].  Further data will be needed to resolve these issues.

\vskip0.2in
\vskip\parskip
\noindent{\bf 4. Conclusions}
\vskip0.1in

We have analyzed data from the MAX experiment in the context
of cold dark matter models of structure formation.
An important feature of our analysis was the incorporation
of the strong anisotropy due to the experimental chopping strategy
in our theoretical autocorrelation function $\Cth$ for the GUM scan.
If we assume that all of the structure in the GUM data can be attributed to
primordial microwave background fluctuations, then the data are consistent
with CDM models normalized by the COBE results at the 95\% confidence level.
However, the GUM results are marginally inconsistent with the MuP results for
our preferred prior distribution and low $Q$ cutoff.
This could be explained if there is a small amount of foreground contamination
in the GUM data, and/or the two-component fit to the MuP data removed
some of the actual CMB signal.

\bigskip

We are extremely grateful to Naoshi Sugiyama for providing us with
CDM radiation power spectra, and to the MAX group for providing us with
their data.  We would also like to thank
Mark Devlin, Todd Gaier, Phil Lubin, Paul Richards, and especially
Josh Gundersen and Peter Meinhold for many invaluable discussions.  This work
was begun while M.S.~was a visitor at the Center for Particle Astrophysics;
he thanks Bernard Sadoulet and Joe Silk for their kind hospitality.
E.B. acknowledges the support of an NSF graduate fellowship.
This work was supported in part by NSF Grant Nos.~PHY--91--16964
and AST--91--20005.

\vskip1in

\centerline{
\vbox{ \offinterlineskip
\halign {\vrule#& \hfil#\hfil& \vrule#& \hfil#\hfil& \vrule#& \hfil#\hfil&
\vrule#& \hfil#\hfil& \vrule#\cr
\noalign{\hrule}
height2pt&\omit&&\omit&&\omit&&\omit&\cr
&\quad$\Ob$\quad&&\quad 95\% CL\ \ GUM\quad&&
 \quad 95\% CL\ \ MuP\quad&&\quad$C_{\rm th}(0)$\quad&\cr
height2pt&\omit&&\omit&&\omit&&\omit&\cr \noalign{\hrule}
height2pt&\omit&&\omit&&\omit&&\omit&\cr
& 0.01&& 20, 40&& 5, 17&& 14.7&\cr
height4pt&\omit&&\omit&&\omit&&\omit&\cr
& 0.03&& 19, 37&& 4, 16&& 17.5&\cr
height4pt&\omit&&\omit&&\omit&&\omit&\cr
& 0.06&& 17, 34&& 4, 15&& 20.8&\cr
height4pt&\omit&&\omit&&\omit&&\omit&\cr
& 0.10&& 15, 31&& 3, 13&& 26.0&\cr
height2pt&\omit&&\omit&&\omit&&\omit&\cr \noalign{\hrule} }} }
\noindent Table 1: The 95\% confidence level lower and upper limits
(90\% enclosed) for $Q\ (\mu$K), from the MAX-GUM and MAX-MuP
scans for a range of $\Ob$, assuming $f(Q)=\log Q$ (see text).
We also give $\Cth(0)$ to show the scaling of the CDM predictions with $\Ob$.

\vfill\eject

\noindent{\bf References}
\vskip0.1in
\frenchspacing
\parindent=0truept \leftskip=0.8truecm \rightskip=0truecm
\everypar{\hangindent=\parindent}

\item{[1]} J.O.~Gundersen, A.C.~Clapp, M.~Devlin, W.~Holmes, M.L.~Fischer,
P.R.~Meinhold, A.E.~Lange, P.M.~Lubin, P.L.~Richards, and G.F.~Smoot,
Astrophys. J. {\bf 413}, L1 (1993).

\item{[2]} M.~Devlin, et al., in Proc. NAS Colloquium on Physical Cosmology
(Irvine), in press.

\item{[3]} P.R.~Meinhold, A.C.~Clapp, D.~Cottingham, M.~Devlin, M.L.~Fischer,
J.O.~Gundersen, W.~Holmes, A.E.~Lange, P.M.~Lubin, P.L.~Richards,
and G.F.~Smoot, Astrophys. J. {\bf 409}, L1 (1993).

\item{[4]} M.~White, L.M.~Krauss, and J.~Silk, Astrophys. J., in press,
astro-ph/9303009.

\item{[5]} G.F.~Smoot, et al., Astrophys. J. {\bf 396}, L1 (1992).

\item{[6]} E.L.~Wright, G.F.~Smoot, A.~Kogut, G.~Hinshaw,  L.~Tenorio,
C.~Lineweaver, C.L.~Bennett, and P.M.~Lubin, COBE Report No.~93-06,
Astrophys. J., in press.

\item{[7]} N.~Sugiyama and N.~Gouda, Prog. Theor. Phys.~{\bf 88}, 803 (1992).

\item{[8]} J.R.~Bond and G.~Efstathiou, Mon. Not. R. Astron. Soc. {\bf 226},
655 (1987).

\item{[9]} J.R.~Bond, G.~Efstathiou, P.M.~Lubin, and P.R.~Meinhold,
Phys. Rev. Lett.~{\bf 66}, 2179 (1991).

\item{[10]} T.~Gaier, J.~Schuster, J.O.~Gundersen, T.~Koch, P.R.~Meinhold,
M.~Seiffert, and P.M.~Lubin, Astrophys. J. {\bf 398}, L1 (1992).

\item{[11]} E.T.~Bunn, M.~White, M.~Srednicki, and D.~Scott,
submitted to Astrophys. J., astro-ph/9308021.

\item{[12]} J.~Gundersen and P.~Meinhold, private communication.

\item{[13]} J.O.~Berger, Statistical decision theory and Bayesian analysis,
2nd ed., (Springer-Verlag, New York, 1985).

\item{[14]} D.~Scott,  M.~Srednicki, and M.~White, submitted to Astrophys. J.,
astro-ph/9305030.

\item{[15]} P.~Meinhold, private communication.

\nonfrenchspacing
\vfill\eject
\leftskip=0truecm
\noindent{\bf Figure Captions}
\vskip0.1in
Fig.~1~~The MAX-GUM scan pattern.  The X's indicate the locations where
the data were binned (into 15 positions along 11 scans), in a coordinate
system where Gamma Ursae Minoris (GUM) is at the origin, and the units are
degrees on the sky.  The scan and chop direction point either toward
(for $y<0$) or away (for $y>0$) from the origin.  For comparison,
the circle has a radius equal to the Gaussian width $\sigma$ of the beam,
and the vertical line shows the size of the peak-to-peak beam chop.

Fig.~2~~An illustrative plot of the theoretical autocorrelation function,
$\Cth(\hat{n}_1,\hat{y};\hat{n}_2,\hat{y})$, where $\hat{n}_1$ points towards
(0,0) (at the back of the plot) and $\hat{n}_2$ points towards $(x,y)$;
$x$ and $y$ are given in degrees on the sky.
Note the strong anisotropy.  For this plot only, the beam is always chopped
in the positive $y$ direction.
$\Cth$ was computed assuming a CDM model with $\Ob=0.06$.

Fig.~3~~The likelihood function ${\cal L}(Q)\,df/dQ$ vs.~$Q\ (\mu$K), for the
MuP and GUM scans of MAX, assuming a CDM model with
$\Ob=0.10$ (long dashed), $0.06$ (short dashed), $0.03$ (dotted), $0.01$
(solid), and using the prior distribution $f(Q)=\log Q$ (see text).
Also shown are the likelihoods for $\Ob=0.01$ with $f(Q)=Q$ and $f(Q)=Q^2$
(the extra solid curves furthest right for each scan).
The two vertical lines bracket the COBE preferred value $(\pm1\sigma)$.

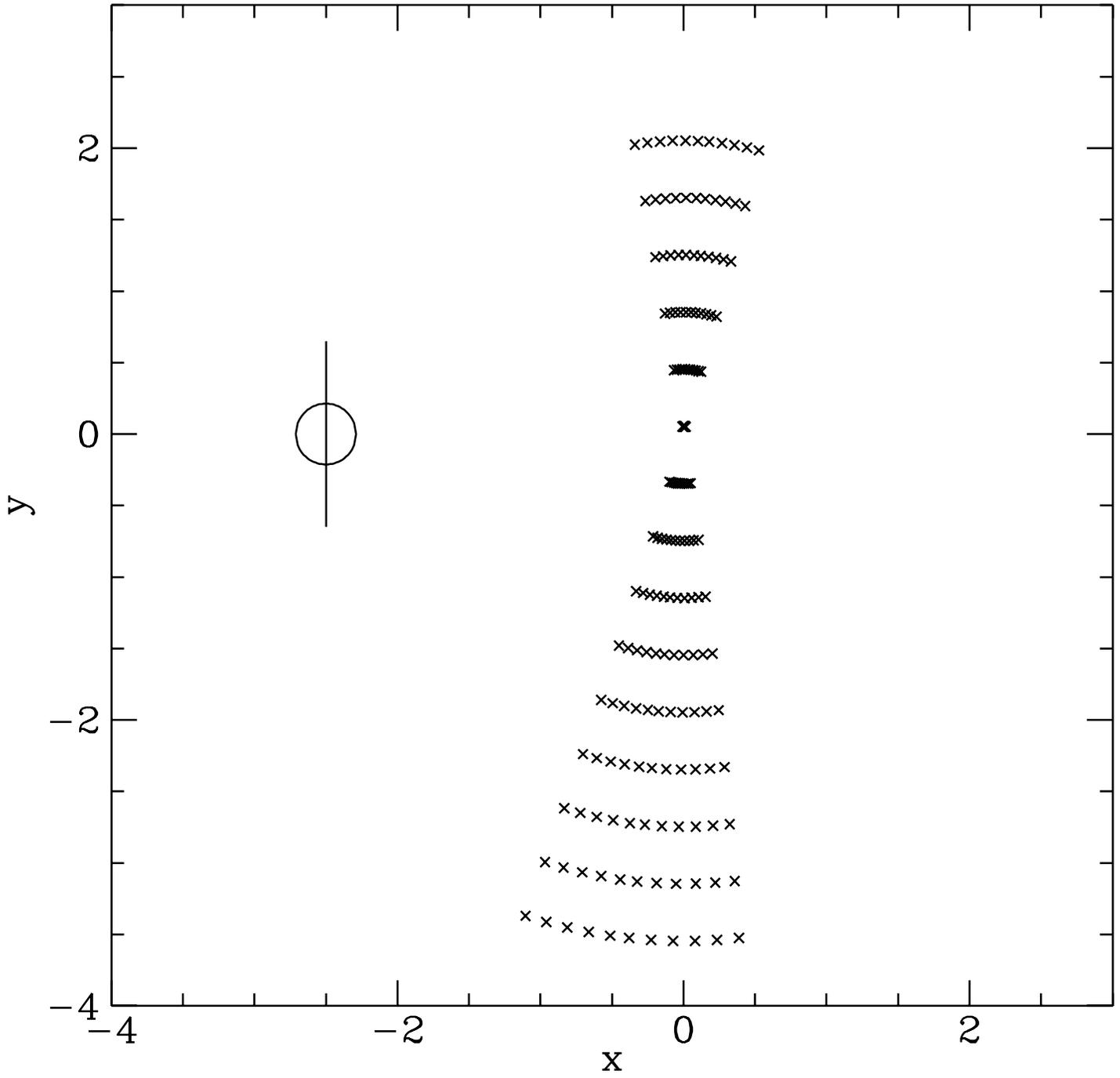
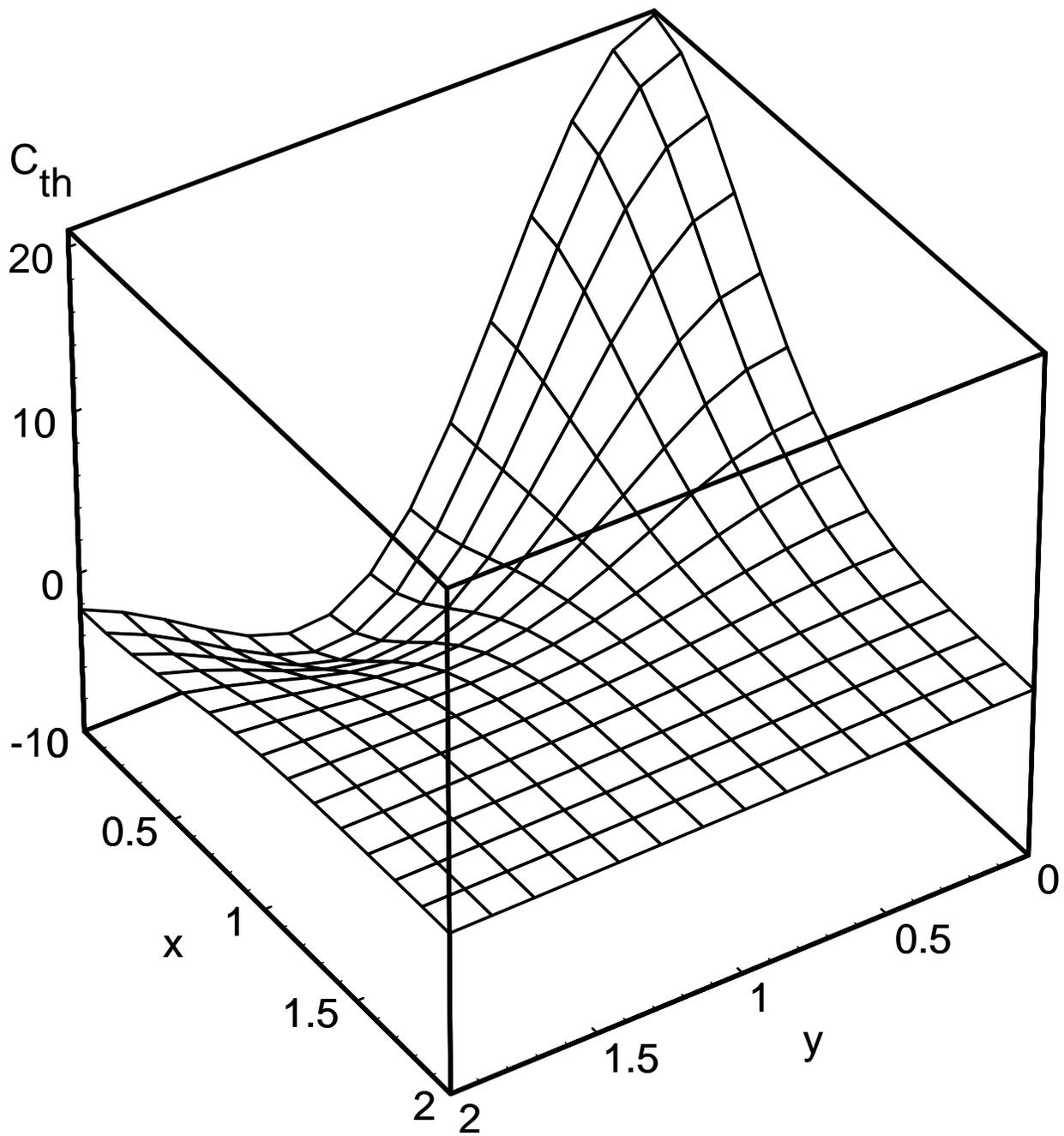
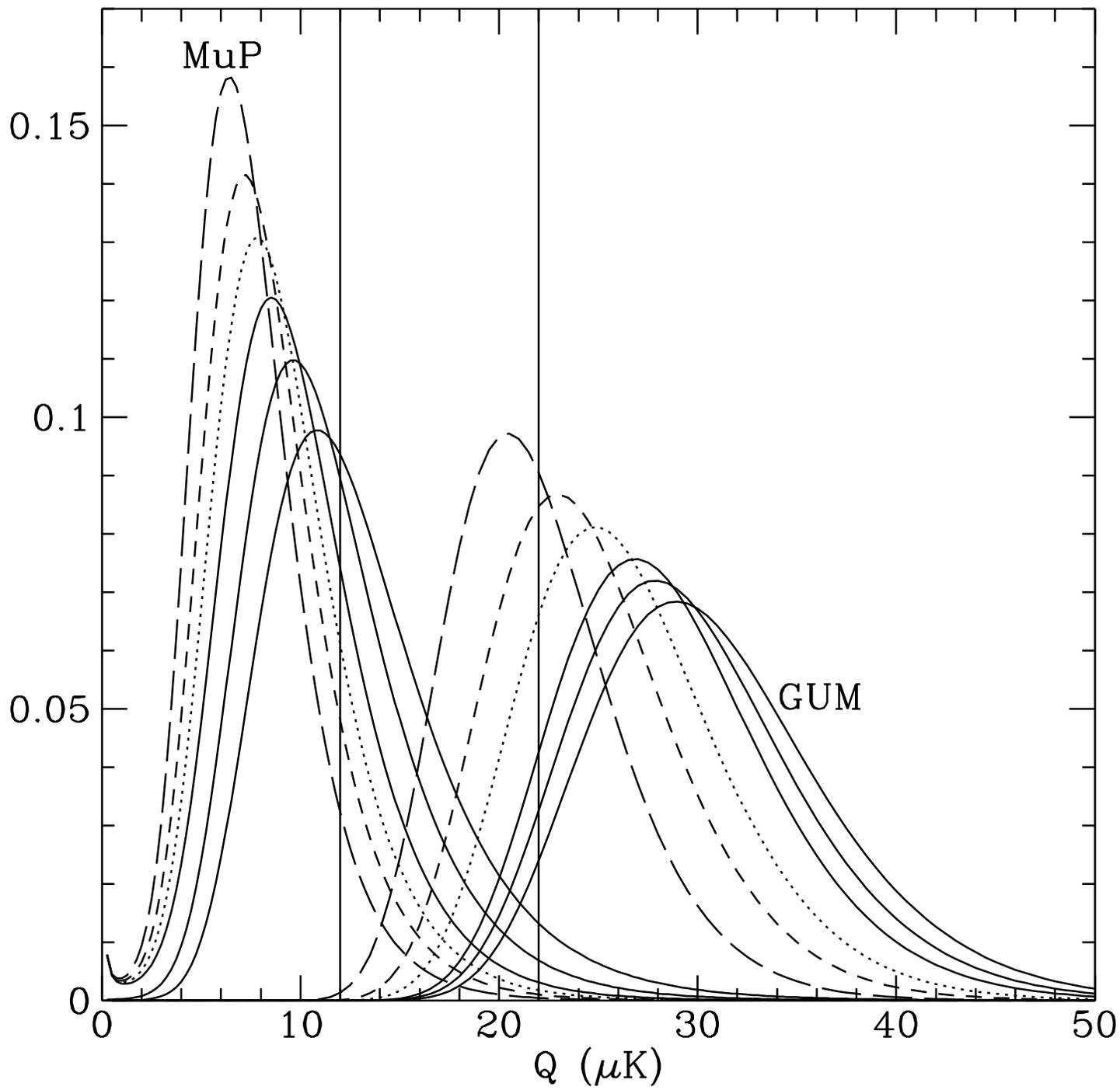
\end